\documentclass[12pt]{article}
\usepackage{latexsym}
\usepackage{amsmath}
\usepackage{amsfonts}
\usepackage{amssymb}
\usepackage{amscd}
\usepackage{fancybox, accents}
\usepackage{cite}
\usepackage{amsmath,amsfonts,amsbsy}
\usepackage{pstricks,pst-node}
\usepackage[small,bf,hang]{caption2}

\usepackage[vcentermath]{youngtab}

\usepackage{float}

\psset{unit=1.3cm,linewidth=.5pt,radius=.2}  

\usepackage[vcentermath]{youngtab}          
\usepackage{enumitem}                       
\usepackage{multirow}                     
\usepackage{float}                          
\usepackage{lscape}                         
\usepackage{bm}

\def\hybrid{\topmargin -20pt    \oddsidemargin 0pt
        \headheight 0pt \headsep 0pt
        \textwidth 6.25in       
        \textheight 9.5in       
        \marginparwidth .875in
        \parskip 5pt plus 1pt   \jot = 1.5ex}

\hybrid

\newcommand{\cI}{{\cal I}}
\newcommand{\cJ}{{\cal J}}
\newcommand{\cK}{{\cal K}}
\newcommand{\cL}{{\cal L}}

\newcommand{\cO}{{\cal O}}
\newcommand{\cP}{{\cal P}}
\newcommand{\cQ}{{\cal Q}}

\newcommand{\cV}{{\cal V}}

\newcommand{\beq}{\begin{equation}}
\newcommand{\eeq}{\end{equation}}
\newcommand{\bi}{\begin{itemize}}
\newcommand{\ei}{\end{itemize}}
\newcommand{\bea}{\begin{eqnarray}}
\newcommand{\eea}{\end{eqnarray}}
\newcommand{\ba}{\begin{array}}
\newcommand{\ea}{\end{array}}
\newcommand{\bt}{\begin{tabular}}
\newcommand{\et}{\end{tabular}}
\newcommand{\bc}{\begin{center}}
\newcommand{\ec}{\end{center}}

\def\theequation{\arabic{section}.\arabic{equation}}

\newcommand{\ft}[2]{{\textstyle {\frac{#1}{#2}} }}

\begin{document}

\begin{titlepage}
\begin{center}

\hfill UG-08-07 \\
\hfill UB-ECM-PF-08/11 \\

\vskip 2cm

{\Large \bf Multiple Membranes 
from Gauged Supergravity
\\[0.2cm]}

\vskip 2cm

{\bf Eric A.~Bergshoeff\,$^1$, Mees de Roo\,$^1$, Olaf Hohm\,$^1$ and Diederik Roest\,$^2$} \\

\vskip 25pt

{\em $^1$ \hskip -.1truecm Centre for Theoretical Physics, University of Groningen, \\
Nijenborgh 4, 9747 AG Groningen, The Netherlands \vskip 5pt }

{email: {\tt E.A.Bergshoeff@rug.nl, M.de.Roo@rug.nl, O.Hohm@rug.nl}} \\

\vskip 15pt

{\em $^2$ \hskip -.1truecm Departament Estructura i Constituents de la Materia \\
    \& Institut de Ci\`{e}ncies del Cosmos, \\
Diagonal 647, 08028 Barcelona, Spain \vskip 5pt }

{email: {\tt droest@ecm.ub.es}} \\

\vskip 0.8cm

\end{center}

\vskip 1cm

\begin{center} {\bf ABSTRACT}\\[3ex]

\begin{minipage}{13cm}
\small Starting from gauged ${\cal N}=8$ supergravity in three
dimensions we construct actions for multiple membranes by taking the
limit to global supersymmetry for different choices of the embedding
tensor. This
provides a general framework that reproduces many recent results on
multiple membrane actions as well as generalisations thereof. As
examples we discuss conformal (non-conformal) gaugings leading to
multiple M2-branes (D2-branes) and massive deformations of these
systems.

\end{minipage}

\end{center}

\noindent

\vfill


June 2008

\end{titlepage}

\tableofcontents

\section{Introduction} \setcounter{equation}{0}

Recently, there has been a lot of activity in constructing actions
for multiple M2-branes. This development was spurred by a series of
papers by Bagger and Lambert \cite{Bagger:2006sk, Bagger:2007jr,
Bagger:2007vi} and Gustavsson
\cite{Gustavsson:2007vu,Gustavsson:2008dy} (following earlier work
of \cite{Schwarz:2004yj,Basu:2004ed}) who made a proposal for a
three-dimensional action describing multiple M2-branes. This action
is an ${\cal N}=8$ superconformal Chern-Simons gauge theory.

It turns out that the original proposal of \cite{Bagger:2006sk,
Bagger:2007jr, Bagger:2007vi,Gustavsson:2007vu,Gustavsson:2008dy} is
rather restrictive. The presence of a so-called fundamental identity
leads to a basically unique solution with $SO(4)$ gauge group (or
direct sums thereof) \cite{Papadopoulos:2008sk,Gauntlett:2008uf}
that describes a system of two M2-branes on an
orbifold
\cite{Lambert:2008et,Distler:2008mk}. To
describe more general M2-brane systems  an extension of the original
proposal is needed and several extensions have  been considered. A
possibility is to consider supersymmetric gauge theories without a
Lagrangian \cite{Gran:2008vi}. Also massive extensions breaking the
conformal invariance have been constructed
\cite{Gomis:2008cv,Hosomichi:2008qk,Song:2008bi}. More recently,
 new extensions to arbitrary gauge groups of the
Bagger-Lambert theory have been proposed that make use of an invariant metric that is
not positive definite
\cite{Gomis:2008uv,Benvenuti:2008bt,Ho:2008ei}. This has the
potentially troublesome feature that it introduces ghosts, an issue
which has been addressed in \cite{Bandres:2008kj, Gomis:2008be,
Ezhuthachan:2008ch}. In addition, Chern-Simons theories with less
supersymmetries  in the context of
M2-branes have been considered \cite{Aharony:2008ug, Benna:2008zy}.
For other related work
on multiple M2-branes, see \cite{Mukhi:2008ux}.

The recent interest in multiple membranes deals with the properties
of globally ${\cal N}=8$ supersymmetric gauge theories in three
dimensions. Independent of this, a lot is  known about the
construction of {\sl locally} ${\cal N}=8$ supersymmetric  theories
in three dimensions. There are a few parallel developments in
constructing theories with global versus local supersymmetry. For
instance, one issue with the construction of an ${\cal N}=8$
supersymmetric gauge theory in three dimensions is the origin of the
gauge fields. To describe M2-brane actions one needs to work with
the maximum number $8N$ of scalar kinetic terms and there is no room
for a vector field kinetic term. The way out was given in
\cite{Schwarz:2004yj}. The vector fields needed for the gauging only
occur inside the covariant derivatives and via a Chern-Simons term
but do not have a kinetic term. Their field equations lead to a
duality relation between the vectors and the scalars such that no
new degrees of freedom are introduced. Precisely the same issue was
encountered in the construction of gauged supergravity in three
dimensions \cite{Nicolai:2000sc,Nicolai:2001sv,Nicolai:2001ac}. For
instance, in ${\cal N}=8$ supergravity all bosonic degrees of
freedom are described by scalars parametrizing the coset $SO(8,N) /
(SO(8)\times SO(N))$, and there are no vector fields left to perform
the gauging. The resolution proposed in
\cite{Nicolai:2000sc,Nicolai:2001sv} is the same as in the globally
supersymmetric case: the vector fields only occur via covariant
derivatives and a Chern-Simons term.

A noteworthy feature of gauged supergravities in three dimensions is
that it suffices to restrict oneselves to theories in which the
Yang-Mills gauge fields only occur via a Chern-Simons term without a
separate kinetic term. This is due to the existence of a non-Abelian
duality which states that any Yang-Mills theory in three dimensions
can be re-interpreted as the sum of kinetic terms for scalar fields
and a $B\wedge F(A)$ Chern-Simons gauge theory (containing two
distinct vector fields $A$ and $B$) based on a non-semi-simple Lie
algebra \cite{Nicolai:2003bp, deWit:2003ja}.  It is via such
Chern-Simons terms that we recover, after applying the non-Abelian
duality, results for multiple D2-brane actions as well. We will also
encounter Chern-Simons gauge theories of the type $A\wedge F(A)$,
which are topologically massive gauge theories.

The construction of \cite{Nicolai:2000sc, Nicolai:2001sv} classifies
the most general gaugings in supergravity, which are encoded in the
`embedding tensor'. The role of this tensor is to specify which
subgroup of the global symmetry group is gauged and which vectors
are needed to perform this gauging. Originally this technique was
developed to construct maximal supergravity in three dimensions and
was later applied to the ${\cal N}=8$ case
\cite{Nicolai:2001ac,deWit:2003ja} and in higher dimensions as
well\cite{deWit:2002vt, deWit:2003hr, deWit:2004nw, deWit:2005hv,
 Samtleben:2005bp,Schon:2006kz,deWit:2007mt,Bergshoeff:2007ef}.
The same technique can be applied to supersymmetric gauge theories.
This has been done to construct the gaugings of ${\cal N}=2$
supersymmetry in four dimensions \cite{deVroome:2007zd} and, more
recently, to reconstruct \cite{Bergshoeff:2008cz} the supersymmetric
gauge theory  of \cite{Bagger:2006sk,
Bagger:2007jr,Bagger:2007vi,Gustavsson:2007vu,Gustavsson:2008dy}. In
the latter case the embedding tensor is a 4-index anti-symmetric
tensor of $SO(N)$ that coincides with the `structure constants' of
the three-algebra occurring in the construction of
\cite{Bagger:2006sk, Bagger:2007jr,
Bagger:2007vi,Gustavsson:2007vu,Gustavsson:2008dy}.

In contrast to the supersymmetric gauge theory  with the unique
$SO(4)$ gauge group of \cite{Bagger:2006sk, Bagger:2007jr,
Bagger:2007vi,Gustavsson:2007vu,Gustavsson:2008dy}, in supergravity
a wide variety of gaugings is possible. In particular, one can embed
the gauge group into the {\sl non-compact group} $SO(8,N)$ whereas
only subgroups of the {\sl compact} $SO(N)$ group are gauged in
\cite{Bagger:2006sk, Bagger:2007jr,
Bagger:2007vi,Gustavsson:2007vu,Gustavsson:2008dy}. In this work we
want to investigate the relation between the two types of theories
and their gaugings. In particular we want to address the following
question: starting from ${\cal N}=8$ gauged supergravity in three
dimensions, can we take the limit of global supersymmetry and if so,
does this lead to known and/or new supersymmetric gauge theories
describing multiple branes?

In order to answer this question we have organised this paper as
follows. In section 2 we will first write down the ${\cal N}=8$
supergravity theory and next consider the limit to global
supersymmetry. Furthermore, we will present the general result for
the globally supersymmetric theory. In  section 3 we focus on the
separate deformations and discuss their interpretation in terms of
multiple branes. Our conclusions are presented in section 4.
Finally, appendix A contains our conventions and useful formulae for
the $SO(8,N)$ structure of supergravity.

\section{Gauged ${\cal N} = 8$ supergravity and its global limit} \setcounter{equation}{0}

\subsection{The Lagrangian and the embedding tensor}

We start by reviewing ${\cal N}=8$ gauged supergravity in $D=3$
\cite{Nicolai:2001ac,deWit:2003ja}. For an overview of our
conventions see appendix A.

The ${\cal N}=8$ supergravity multiplet consists of the metric
$g_{\mu\nu}$ and $8$ gravitini $\psi_{\mu}^{A}$. All these fields
are topological and do not describe physical degrees of freedom.
Therefore, all local degrees of freedom reside in scalars and
Majorana spinor fields. In the case of $N$ matter multiplets, there
are $8N$ scalars $X^{aI}$, $a=1,\ldots,N$, parameterizing the coset
space $SO(8,N)/(SO(8)\times SO(N))$, and $8N$ spinors denoted by
$\chi^{\dot{A}a}$. The coset dynamics of the scalar fields is
expressed in terms of the group-valued matrix $L(x) \in SO(8,N)$. It
can be parameterized in terms of scalars in the following way
 \bea\label{scalar}
  L(x) \ = \ \exp\left(X^{Ia}(x)t^{Ia}\right)\;,
 \eea
where $\{t^{IJ},t^{ab}\}$ and $\{t^{Ia}\}$ denote the compact and
non-compact generators of $\frak{so}(8,N)$, c.f.~the appendix. To be
more precise,  we have gauge-fixed  the local $SO(8)\times SO(N)$
symmetry by setting the compact part of $L$ to zero.

In order to gauge a certain subgroup $G_0$ of the (rigid) duality
group $SO(8,N)$ one introduces gauge-covariant derivatives in the
definition of the Maurer-Cartan forms as follows:
 \bea\label{maurer}
  L^{-1}\left(\partial_{\mu}+\Theta_{\alpha\beta}A_{\mu}{}^{\alpha}t^{\beta}\right)L
   \ =: \ \ft12 {\cal Q}_{\mu}{}^{IJ}t^{IJ}+\ft12 {\cal
   Q}_{\mu}{}^{ab}t^{ab}+{\cal P}_{\mu}{}^{Ia}t^{Ia}\;.
 \eea
Here $t^{\alpha}$ denote the generators of $SO(8,N)$,
$\alpha,\beta,\ldots =1,\ldots,\ft12(N+8)(N+7)$, with structure
constants $f^{\alpha\beta}{}_{\gamma}$. Furthermore, we have
introduced gauge fields $A_\mu{}^\alpha$ in the adjoint
representation of $\frak{so}(8,N)$ and the symmetric embedding
tensor $\Theta_{\alpha\beta}$ \cite{Nicolai:2001ac}. The latter
encodes the embedding of the gauge group $G_0$ into the global
symmetry group $SO(8,N)$. To be more precise, the generators of
$G_0$ are given by
 \bea\label{gaugegen}
  X_{\alpha}=\Theta_{\alpha\beta}t^{\beta}\;,
 \eea
and so the embedding tensor singles out those generators that span
the gauge group. In particular, the dimension of the gauge group is
determined by the rank of $\Theta_{\alpha\beta}$.

The gauged supergravity Lagrangian is completely determined by the
embedding tensor and given by \cite{Nicolai:2001ac}
 \begin{eqnarray}\label{sugra}
  {\cal L} &=&  -\tfrac{1}{2} \kappa^{-2}e\,R+
  \varepsilon^{\mu\nu\rho}\bar{\psi}_{\mu}^{A}D_{\nu}\psi_{\rho}^{A}
  +\tfrac{1}{2} \kappa^{-2}e\,{\cal P}_{\mu}{}^{Ia}{\cal
  P}{}^{\mu\hspace{0.1em}Ia}-ie\,\bar{\chi}^{\dot{A}a}\gamma^{\mu}D_{\mu}\chi^{\dot{A}a}\\
  \nonumber
  &&-\ft12
  \Theta_{\alpha\beta}\varepsilon^{\mu\nu\rho}A_{\mu}{}^{\alpha}\left(\partial_{\nu}A_{\rho}{}^{\beta}+\ft13
  \Theta_{\gamma\delta}f^{\beta\delta}{}_{\epsilon}A_{\nu}{}^{\gamma}A_{\rho}{}^{\epsilon}\right)
  - e\,{\cal
  P}_{\mu}{}^{Ia}\bar{\chi}^{\dot{A}a}\Gamma^{I}_{A\dot{A}}\gamma^{\nu}\gamma^{\mu}\psi_{\nu}^{A}\\
  \nonumber
  && + \kappa^{-2}e\,A_1^{AB}\bar{\psi}_{\mu}^{A}\gamma^{\mu\nu}\psi_{\nu}^{B}+2i\kappa^{-2}e\,A_2^{A,\dot{A}a}\bar{\chi}^{\dot{A}a}
  \gamma^{\mu}\psi_{\mu}^{A}+\kappa^{-2}e\,A_3^{\dot{A}a,\dot{B}b}\bar{\chi}^{\dot{A}a}\chi^{\dot{B}b}-\kappa^{-6}e\,V\;,
 \end{eqnarray}
where $\kappa$ is the square root of Newton's constant with mass
dimension\footnote{The mass dimensions of
 $\{g_{\mu\nu}\,, \psi_\mu^{A}\,, A_\mu{}^\alpha \,, \chi^{{\dot A} a}\,, X^{Ia} \}$ are
 given by $\{0\,, 1\,, 1\,, 1\,, 0\}$.}
$-\ft12$ and $\gamma^{\mu}$ and $\Gamma^{I}_{A\dot{A}}$
are gamma matrices of $SO(1,2)$ and $SO(8)$, respectively. Furthermore, the
covariant derivatives $D_{\mu}$ on the spinors are given by
 \bea
 \begin{split}
  D_{\mu}\psi_{\nu}^{A} \ &= \
  \nabla_{\mu}\psi_{\nu}^{A}
  +\ft14\cQ_{\mu}^{IJ}\Gamma^{IJ}_{AB}\psi_{\nu}^{B}\;, \\
  D_{\mu}\chi^{\dot{A}a} \ &= \
  \nabla_{\mu}\chi^{\dot{A}a}
  +\ft14
  \cQ_{\mu}^{IJ}\Gamma^{IJ}_{\dot{A}\dot{B}}\chi^{\dot{B}a}
  +\cQ_{\mu}^{ab}\chi^{\dot{A}b}\;,
 \end{split}
 \eea
and contain the (composite) $SO(8)\times SO(N)$ connections defined
in (\ref{maurer}). Finally, the scalar-dependent Yukawa couplings given by $A_{1,2,3}$ and the
scalar potential $V$ are completely determined by the embedding
tensor via the so-called T-tensor
 \bea
  T_{\underaccent{\bar}{\alpha},\underaccent{\bar}{\beta}} \ = \ \Theta_{\alpha\beta}{\cal
  V}^{\alpha}{}_{\underaccent{\bar}{\alpha}}{\cal
  V}^{\beta}{}_{\underaccent{\bar}{\beta}}\;,
 \eea
where $\underline{\alpha},\underline{\beta},\ldots$ are flat indices
corresponding to the local $SO(8)\times SO(N)$ action. Here ${\cal
V}$ is the adjoint $SO(8,N)$ matrix, defined through
 \bea\label{adjoint}
  L^{-1}t^{\alpha}L \ = \ {\cal
  V}^{\alpha}{}_{\underaccent{\bar}{\alpha}}t^{\underaccent{\bar}{\alpha}} \ = \
  \ft12\cV^{\alpha}{}_{IJ}t^{IJ}+\ft12 \cV^{\alpha}{}_{ab}t^{ab}+
  \cV^{\alpha}{}_{Ia}t^{Ia}\;.
 \eea
In terms of the T-tensor the Yukawa couplings read
 \begin{eqnarray}\label{A123}
  A_1^{AB} &=&
  -\delta^{AB}\theta-\tfrac{1}{48}\Gamma^{IJKL}_{AB}T_ {IJ,KL}\;,
  \qquad
  A_2^{A,\dot{A}a} \ = \
  -\tfrac{1}{12}\Gamma^{IJK}_{A\dot{A}}T_{IJ,Ka}\;, \\ \nonumber
  A_3^{\dot{A}a,\dot{B}b} &=&
  2\delta^{\dot{A}\dot{B}}\delta^{ab}\theta
  +\tfrac{1}{48}\delta^{ab}\Gamma^{IJKL}_{\dot{A}\dot{B}}T_{IJ,KL}
  +\tfrac{1}{2}\Gamma^{IJ}_{\dot{A}\dot{B}}T_{IJ,ab}\;,
 \end{eqnarray}
where
$\theta=\tfrac{2}{(N+8)(N+7)}\eta^{\alpha\beta}\Theta_{\alpha\beta}$
denotes the trace of the embedding tensor with respect to the
Cartan-Killing form $\eta^{\alpha\beta}$ of $SO(8,N)$. The scalar
potential $V$ is given by
 \bea
  V \ = \
  -\tfrac{1}{2}\left(A_1^{AB}A_1^{AB}-\tfrac{1}{2}A_2^{A,\dot{A}a}A_2^{A,\dot{A}a}\right)\;.
 \eea

The local supersymmetry transformations leaving (\ref{sugra})
invariant are given by
 \begin{eqnarray}\label{susy}
  \delta_{\epsilon}e_{\mu}{}^{r} &=&
  i\kappa\,\epsilon^{A}\gamma^{r}\psi_{\mu}^{A}\;, \qquad
  \delta_{\epsilon}\psi_{\mu}^{A} \ = \ \kappa^{-1}\,
  D_{\mu}\epsilon^{A}+i \kappa^{-3}\,A_1^{AB}\gamma_{\mu}\epsilon^{B}\;, \\  \nonumber
    \delta_{\epsilon}A_{\mu}{}^{\alpha} &=& -\tfrac{1}{2} \kappa^{-1} \,{\cal
  V}^{\alpha}{}_{IJ}\bar{\epsilon}^{A}\Gamma^{IJ}_{AB}\psi_{\mu}^{B}
  +i\kappa^{-1} \,{\cal
  V}^{\alpha}{}_{Ia}\bar{\epsilon}^{A}\Gamma^{I}_{A\dot{A}}\gamma_{\mu}\chi^{\dot{A}a}\;, \\ \nonumber
  \delta_{\epsilon}\chi^{\dot{A}a} \ &=& \ \ft12 i \kappa^{-1} \,
  \Gamma^{I}_{A\dot{A}}\gamma^{\mu}\epsilon^{A}{\cal
  P}_{\mu}{}^{Ia}+\kappa^{-3}\,A_2^{A,\dot{A}a}\epsilon^{A}\;,
  \qquad
    L^{-1}\delta_{\epsilon}L = \kappa\,\bar{\epsilon}^{A}\Gamma^{I}_{A\dot{A}}\chi^{\dot{A}a}t^{Ia}\;,
    \nonumber
 \end{eqnarray}
where we assign mass dimension $-\ft12$ to the supersymmetry
parameter $\epsilon$.

However, consistency of the gauged supergravity theory requires linear and quadratic constraints on the
embedding tensor. First of all, gauge invariance of (\ref{sugra})
requires invariance of the embedding tensor $\Theta_{\alpha\beta}$
under the adjoint action of the gauge group generators $X_{\alpha}$.
This implies the quadratic constraint
 \bea\label{quadconstr}
  {\cal Q}_{\alpha,\beta\gamma} \ \equiv \
  \Theta_{\alpha\delta}\Theta_{\epsilon(\beta}f^{\delta\epsilon}{}_{\gamma)}
  \ = \ 0\;,
 \eea
which is also sufficient for closure of the gauge algebra. Beyond
this quadratic constraint, invariance of (\ref{sugra}) under
supersymmetry requires a linear constraint, which takes a ``duality
covariant'' form. The embedding tensor reads
$\Theta_{\alpha\beta}=\Theta_{[\cI\cJ],[\cK\cL]}$, $\cI,\cJ,\ldots
=1,\ldots,8+N$, where we have introduced adjoint indices for
$SO(8,N)$. Due to its symmetry, a priori it takes values in the
symmetric tensor product
 \bea\label{irreducible}
  \left(\hspace{0.2em}{\small \yng(1,1)}\otimes {\small \yng(1,1)}\hspace{0.2em}\right)_{\rm sym}
  \ = \  {\bf 1} \oplus {\small \yng(1,1,1,1)} \oplus
  {\small \yng(2,2)}\oplus {\small \yng(2)}\;,
 \eea
in which the Young tableaux refer to tensors of the full duality
group $SO(8,N)$. However, supersymmetry restricts the irreducible
representations in (\ref{irreducible}) to a subclass. Specifically,
in the given case it eliminates the irreducible representation
corresponding to the window tableau \cite{deWit:2003ja}. In other
words, the linear constraint can be written as
 \bea \label{linconst}
  \Theta_{\cI\cJ,\cK\cL}  \ = \ \theta \delta_{\cI[\cK}\delta_{\cL]\cJ}
  + 2 f_{\cI\cJ\cK\cL}+h_{[\cK[\cI}\delta_{\cJ]\cL]}\;,
   \eea
where $f_{\cI\cJ\cK\cL}$ is totally antisymmetric and $h_{\cI\cJ}$
symmetric--traceless. For any choice of the embedding tensor
satisfying the quadratic and linear constraints (\ref{quadconstr})
and (\ref{linconst}) one obtains a consistent gauged supergravity,
which is invariant under the supersymmetry transformations
(\ref{susy}).\footnote{We should note that the expressions
(\ref{A123}) for the Yukawa couplings are valid provided the
embedding tensor satisfies the stronger constraint $h_{\cI\cJ}=0$.
We verified that in the presence of $h_{\cI\cJ}$ one can still take
the global limit to be discussed below. For the general formulae see
\cite{deWit:2003ja}.}\label{footnote}

\subsection{The limit to global supersymmetry}\label{global}

We will now discuss the limit to global supersymmetry, i.e.~we
decouple gravity by sending Newton's constant, or its square root
$\kappa$, to zero. We will find that this limit can only be taken provided
a number of
additional constraints is imposed on the embedding tensor.

To take the flat space limit, we have to expand the metric around
Minkowski spacetime according to
\begin{equation}
g_{\mu \nu} \ = \ \eta_{\mu \nu} + \kappa h_{\mu \nu}\,.
\end{equation}
In the limit $\kappa\rightarrow 0$, the spin-2 multiplet
$\{h_{\mu\nu}\,, \psi_\mu^{A}\}$ decouples and can therefore be set
to zero. This restricts the parameters $\xi_\mu$ of general
coordinate transformations and the parameters $\epsilon^A$ of
supersymmetry to those satisfying the equations
$\partial_\mu\,\xi_\nu +
\partial_\nu\,\xi_\mu=\partial_{\mu}\epsilon^A=0$. Thus,
we obtain a globally supersymmetric theory, in which $\epsilon^A$ is
constant. Moreover, in order to obtain a non-singular limit, in
which non-trivial gaugings survive, it turns out to be necessary to
rescale the fields and various components of the embedding tensor
with powers of $\kappa$. Specifically, we redefine the scalar fields
according to
 \begin{equation}
  X^{Ia} \rightarrow \kappa\, X^{Ia}\;,
 \end{equation}
such that their mass dimension is $\ft12$, and we redefine the gauge vectors
depending on their $SO(8) \times SO(N)$ indices according to
 \begin{align}\label{vectorscale}
  A_\mu{}^{ab} \rightarrow A_\mu{}^{ab} \,, \qquad
  A_\mu{}^{aI} \rightarrow \kappa^{-1} A_\mu{}^{aI} \,, \qquad
  A_\mu{}^{IJ} \rightarrow \kappa^{-1} A_\mu{}^{IJ} \,.
 \end{align}
Moreover, we require the following scaling weights for the
components of the embedding tensor,
 \begin{align}
  & \Theta_{ab,cd} \, : \, 0 \,, \qquad
  \Theta_{ab,cI} \,,
  \Theta_{ab,IJ} \, : \, 1 \,, \qquad
   \Theta_{aI,cJ} \,,
  \Theta_{aI,JK} \,,
  \Theta_{IJ,KL} \, : \, 2 \,
  \label{Theta-weights}
 \end{align}
where we indicated the powers of $\kappa$.

Let us now explain the limit and the origin of the different scaling
weights in more detail. First of all, inspection of the
scalar-kinetic terms shows by use of the expansion of the
Maurer-Cartan forms (\ref{PQ}) that the terms of higher order in $X$
will drop out. In other words, the scalar manifold reduces to a flat
space. This can be interpreted as an In\"on\"u-Wigner contraction of
the original coset space, for which one rescales the non-compact
generators by $\bar{t}^{Ia}=\kappa t^{Ia}$ and sends
$\kappa\rightarrow 0$. This leaves the algebra, see
eq.~\eqref{so8N}, intact, except the brackets in the last line,
which become Abelian. Put differently, the Lie algebra reduces to a
semi-direct product between $SO(8)\times SO(N)$ and $8N$
translations. The coset space reduces accordingly to
 \bea
  \frac{\left(SO(8)\times SO(N)\right)\ltimes
  \mathbb{R}^{8N}}{SO(8)\times SO(N)} \cong \mathbb{R}^{8N}\;.
 \eea
Note that the isometry group $ISO(8N)$ of $\mathbb{R}^{8N}$ is much
larger than the expected symmetry group $\left(SO(8) \times  SO(N)
\right) \ltimes \mathbb{R}^{8N}$. However, this symmetry enhancement
only holds for the scalar kinetic terms, and does not extend to the
full theory.

The scaling weights of the gauge fields are uniquely determined by
requiring that the supersymmetry transformations of the vectors in
\eqref{susy} are both non-singular and non-zero in the limit $\kappa
\rightarrow 0$. For instance, one finds
 \begin{align}\label{trrule}
  \delta_\epsilon A_\mu{}^{IJ} = - \bar {\epsilon}^A \Gamma^{IJ}_{AB}\,\psi_\mu^B
  \,.
 \end{align}
One may verify that the global supersymmetry algebra is realized on
these vector fields provided that the following constraints are
satisfied:\footnote{Similarly, it has recently been found that one
can realize the supersymmetry algebra of $D=5, \mathcal{N} = 2$
supergravity on $(D-2)$-forms with vanishing field strengths
\cite{Gomis:2007gb}.}
 \begin{equation}
  \varepsilon^{\mu\nu\rho} \partial_\nu A_\rho{}^{IJ} = 0 \,.
 \end{equation}
We note that the supersymmetry variation of these gauge vectors is
proportional to the gravitino. Therefore they belong to the
topological spin-2 multiplet
$\{g_{\mu\nu}\,,\psi_\mu^{A}\,, A_\mu{}^{IJ}\}$ and we will
henceforth set them to zero. In contrast, we will see below that the
other two representations of gauge vectors in \eqref{vectorscale}
are related to the matter spinors $\chi$ under supersymmetry.
Therefore they belong to the matter multiplets and will make their
appearance in the globally supersymmetric theory.
We also note that the scaling weights (\ref{Theta-weights})
lead to a non-singular limit for the leading Chern-Simons terms,
which otherwise would require certain components of the embedding
tensor to vanish.

Let us now turn to the constraints of the embedding tensor describing
globally supersymmetric
theories. The linear constraints (\ref{linconst}) ensure that the
gauged supergravity action is a consistent starting point. However,
in order to have a well-defined limit further constraints are
required, which we derive by inspecting the Yukawa couplings $A_i$
with $i=1,2,3$.
We first consider the scalar potential. To avoid any
divergent terms, both $A_1$ and $A_2$ have to scale with weight $3$.
Turning to the supersymmetry variation of the gravitino, the right
hand side only vanishes if $A_1$ actually scales with weight $4$.
This has the effect that $A_1$ completely drops out of the theory in
the global limit, as expected. Finally, the scaling weight of $A_3$
has to be $2$, as can be deduced from the relevant term in the
Lagrangian. We thus end up with the following scaling weights for
$A_1, A_2$ and $A_3$\,:
\begin{equation}
A_1\,:\   4 \,, \qquad\qquad\qquad A_2\,:\   3\,, \qquad\qquad\qquad
A_3\,:\ 2\,.
\end{equation}
From the expressions \eqref{A123}, \eqref{TExpan} for $A_1, A_2$ and
$A_3$ we deduce that the above scaling requirements lead to the
following linear constraints on the embedding tensor:
\begin{eqnarray}\label{constraints1}
&& \theta = 0 \,, \qquad \Theta_{ab,IJ} = 0 \,, \qquad
\Theta_{IJ,KL}^- = 0\,, \qquad \Theta_{a[I,JK]} =
0\,,\label{constraints2}
\end{eqnarray}
where $\Theta_{IJ,KL}^-$ denotes the anti-self-dual part of
$\Theta_{IJ,KL}$.

Apart from the constraints (\ref{constraints1}) resulting from the
requirement of a non-singular limit, there is a second source of
linear constraints. This is related to the fact that
 the original linear constraint (\ref{linconst}) of supergravity
 cannot simply be taken over to the globally supersymmetric case
 due to the following reason. The symmetric-traceless
solution $h_{\cI\cJ}$, for instance, in general gives rise to
components of the embedding tensor that scale differently. For
instance, if $h_{\cI\cJ}$ takes non-zero values only in the $SO(N)$
direction, parameterized by a symmetric-traceless $SO(N)$ tensor
$h_{ab}$, one obtains from (\ref{linconst}) the
components\footnote{We thank Hermann Nicolai for discussions on this
point.}
 \bea\label{above}
  \Theta_{ab,cd} \ = \ h_{[a[c}\delta_{d]b]}\;, \qquad
  \Theta_{aI,bJ} \ = \ \ft14 h_{ab}\delta_{IJ}\;.
 \eea
Since according to (\ref{Theta-weights}) we keep the first component
unchanged, while rescaling the second embedding tensor by
$\kappa^2$, the resulting couplings live in different sectors
characterized by embedding tensors of different mass dimension. In
general the supersymmetry will therefore be violated. Thus, in order
to maintain supersymmetry, we have to impose additional linear
constraints, eliminating all solutions of (\ref{linconst}) which
give rise to relations between different components of $\Theta$ with
different scaling weights, like in \eqref{above}. This sets the
singlet $\theta$ to zero, which follows already from
(\ref{constraints1}), as well as the components $f_{IJab}$ of the
4-index anti-symmetric tensor and the components $h_{ab}$ and
$h_{Ia}$ of the symmetric-traceless tensor.

Summarizing, we find from the above considerations that the
only components of the embedding tensor that survive the limit of
global supersymmetry are given by
\begin{equation}\label{linglob}
f_{abcd}\,,\hskip 2truecm
f_{abcI}\,,\hskip 2truecm
f^+_{IJKL}\,, \hskip 2truecm
h_{IJ}\,,
\end{equation}
where $f_{IJKL}^+$ indicates the self-dual part of $f_{IJKL}$.

Finally, we consider the quadratic constraints. One way to derive
these constraints after the rescalings is to consider the gauge
variation of the action before imposing any constraints. For
instance, the Chern-Simons term varies according to
\cite{Bergshoeff:2008cz,Bergshoeff:2008qd}
 \bea\label{CSvar}
  \delta {\cal L}_{\rm CS} \ \sim \ \varepsilon^{\mu\nu\rho}{\cal
  Q}_{\alpha,\beta\gamma}A_{\mu}{}^{\alpha}A_{\nu}{}^{\beta}D_{\rho}\Lambda^{\gamma}\;.
 \eea
Suppose the gauge vectors and their symmetry parameters in
$AAD\Lambda$ will scale with $\kappa^{-r}$ for some $r$, as follows
from (\ref{vectorscale}). Then only those parts of
$\cQ_{\alpha,\beta\gamma}$ will not disappear in the limit
$\kappa\rightarrow 0$, whose dependence on $\kappa$ is $\kappa^s$
with $s \le r$.  For instance, since the $SO(N)$ gauge vectors do
not scale with $\kappa$, only those terms in ${\cal Q}_{ab,cd,ef}$
should be imposed as a constraint that scale with $\kappa^0$. This
in turn implies that in the latter component only the pure $SO(N)$
structure constants enter, while in the full quadratic constraints
of supergravity also the non-compact $f^{aI,bJ}{}_{cd}$ appear.
Similarly, one derives for all other components that the non-trivial
parts of the quadratic constraint tensor involve only the structure
constants corresponding to $SO(N) \ltimes \mathbb{R}^{8N}$. In other
words, denoting these structure constants, i.e.~$f^{ab,cd}{}_{ef}$
and $f^{ab,cI}{}_{dJ}$, collectively by
$\bar{f}^{\alpha\beta}{}_{\gamma}$, the quadratic constraints
imposed by global supersymmetry take formally the same form as in
(\ref{quadconstr}), but with $f$ replaced by $\bar{f}$,
\bea\label{globalquad}
  \cQ^{}_{\alpha,\beta\gamma} \ \equiv \
  \Theta^{}_{\alpha\delta}\Theta^{}_{\epsilon(\beta}\bar{f}^{\delta\epsilon}{}_{\gamma)} \ = \ 0\;.
 \eea
Moreover, since we set $A_{\mu}{}^{IJ}$ and its gauge parameter to
zero, all components of (\ref{globalquad}) whose external indices
take values in the $[IJ]$ direction, need not to be imposed as
constraints. Explicitly one then finds the following non-trivial
components:
 \begin{eqnarray}\label{quadcomp}
  \cQ_{ab,cd,ef} &=&
  \ft12\big(\Theta_{ab,eg}\Theta^{g}{}_{f,cd}-\Theta_{ab,fg}\Theta^{g}{}_{e,cd}
  +\Theta_{ab,cg}\Theta^{g}{}_{d,ef}-\Theta_{ab,dg}\Theta^{g}{}_{c,ef}\big)\;, \\ 
  \cQ_{aI,bJ,cd} &=&
  \ft12\left(\Theta_{aI,c}{}^{g}\Theta_{gd,bJ}-\Theta_{aI,d}{}^{g}\Theta_{gc,bJ}
  -\Theta_{aI,gb}\Theta^{g}{}_{J,cd}\right) \;, \\ 
  \cQ_{ab,cd,eI} &=&
  \ft12\left(\Theta_{ab,eg}\Theta^{g}{}_{I,cd}+\Theta_{ab,gI}\Theta^{g}{}_{e,cd}
  +\Theta_{ab,cg}\Theta^{g}{}_{d,eI}-\Theta_{ab,dg}\Theta^{g}{}_{c,eI}\right)\;,
  \\ 
  \cQ_{aI,bc,de} &=&
  \ft12\left(\Theta_{aI,d}{}^{h}\Theta_{he,bc}-\Theta_{aI,e}{}^{h}\Theta_{hd,bc}
  +\Theta_{aI,b}{}^{h}\Theta_{hc,de}-\Theta_{aI,c}{}^{h}\Theta_{hb,de}\right)\;,
  \\ 
  \cQ_{ab,cI,dJ} &=&
  \Theta_{ab,c}{}^{e}\Theta_{eI,dJ}+\Theta_{ab,d}{}^{e}\Theta_{eJ,cI}\;,
  \\ 
  \cQ_{aI,bJ,cK} &=&
  \Theta_{aI,b}{}^{d}\Theta_{dJ,cK}+\Theta_{aI,c}{}^{d}}\Theta_{dK,bJ\;,
 \end{eqnarray}
where all indices are raised and lowered with the ordinary Kronecker
symbol.

Let us note that in our present analysis the quadratic constraints
have been simplified as compared to supergravity, since we
effectively deal only with gauge groups inside $SO(N) \ltimes
\mathbb{R}^{8N}$. In contrast, the linear constraints are as in
supergravity, but supplemented with further constraints. However,
this does not exclude the possibility that there exist
globally-supersymmetric ${\cal N}=8$ theories that satisfy weaker
constraints, but which cannot be obtained as limits of supergravity
in the given way.

\subsection{The globally supersymmetric ${\cal N}=8$ theory}\label{general}

In this subsection we summarize the resulting globally
supersymmetric action, after taking the limit of ${\cal N}=8$ gauged
supergravity\footnote{We will omit overall $\kappa$-dependences
w.r.t.~the supergravity expressions, as these will drop out.} as
defined in the previous section.

The Lagrangian of the globally supersymmetric theory is given by
 \begin{eqnarray}\label{fieldtheory}
  {\cal L} &=& +\tfrac{1}{2}\,D_{\mu}X^{Ia}D^{\mu}X^{Ia}
  -ie\,\bar{\chi}^{\dot{A}a}\gamma^{\mu}D_{\mu}\chi^{\dot{A}a}
  -\ft12
  \Theta_{\alpha\beta}\varepsilon^{\mu\nu\rho}A_{\mu}{}^{\alpha}\left(\partial_{\nu}A_{\rho}{}^{\beta}
  +\ft13
  \Theta_{\gamma\delta}{\bar f}^{\beta\delta}{}_{\epsilon}A_{\nu}{}^{\gamma}A_{\rho}{}^{\epsilon}\right)
  \notag
  \\
  && + \,A_3^{\dot{A}a,\dot{B}b}\bar{\chi}^{\dot{A}a}\chi^{\dot{B}b} -
  V\;.
 \end{eqnarray}
Here the covariant derivatives of the scalars and fermions are
\begin{align}
& D_{\mu}X^{aI} = \partial_\mu X^{aI} + \Theta_{ab,cd} A_\mu{}^{cd}
 X^{bI} +\ft12 \Theta_{aI,bc} A_\mu{}^{bc}
 + \Theta_{ab,cJ} A_{\mu}{}^{cJ} X^{bI} + \Theta_{aI,bJ} A_\mu{}^{bJ}
 \,, \notag \\
& D_{\mu}\chi^{\dot{A}a} \ = \
  \partial_{\mu}\chi^{\dot{A}a}
  +\cQ_{\mu}{}^{ab}\chi^{\dot{A}b}\,, \qquad
 \mathcal{Q}_\mu{}^{ab} = \ft12 \Theta_{ab,cd} A_{\mu}{}^{cd} +
 \Theta_{ab,cI} A_\mu{}^{cI} \, \label{cov-der-glob}
\end{align}
while the different Yukawa couplings are given by\footnote{See,
however, the provision made in footnote 2.}
 \begin{align} \label{Yukawa-glob}
  A_2^{A,\dot{A}a} = - \tfrac{1}{12} \Gamma_{A \dot{A}}^{IJK}
 ( &  -\Theta_{ab,cd}X^{b}{}_{K}X^{c}{}_{I}X^{d}{}_{J}
  -\Theta_{aK,bc}X^{b}{}_{I}X^{c}{}_{J} + 2\Theta_{ab,c[I}X^{c}{}_{J]}X^{b}{}_{K}
+ \notag \\
  & + 2\Theta_{aK,b[I}X^{b}{}_{J]}
   +\Theta_{IJ,KL}X^{L}{}_{a} ) \,, \notag \\
  A_3^{\dot{A}a,\dot{B}b} =
   + \tfrac{1}{48} \delta^{ab} & \Gamma^{IJKL}_{\dot{A}\dot{B}} \Theta_{IJ,KL}
   + \tfrac{1}{2} \Gamma^{IJ}_{\dot{A}\dot{B}} (
   -\Theta_{ab,cd}X^{c}{}_{I}X^{d}{}_{J} + 2 \Theta_{ab,c[I}X^{c}{}_{J]}) \;,
 \end{align}
The scalar potential is positive definite in the global case and
reads
 \begin{align}
  V \ = \ \tfrac{1}{4} A_2^{A,\dot{A}a} A_2^{A,\dot{A}a} \;.
 \end{align}

The Lagrangian is invariant under the following global ${\cal N} =
8$ supersymmetry transformations:
  \begin{equation}
   \begin{split}
    \delta_{\epsilon}X^{Ia} &= \bar{\epsilon}^{A}\Gamma^{I}_{A\dot{A}}\chi^{\dot{A}a}\;,
    \qquad
    \delta_{\epsilon}\chi^{\dot{A}a} \ = \ \ft12 i
    \Gamma^{I}_{A\dot{A}}\gamma^{\mu}\epsilon^{A}D_{\mu}X^{Ia} + A_2^{A,\dot{A}a}\epsilon^{A}\;, \\
    \delta_\epsilon A_\mu^{ab} &=
    -2i\,\bar{\epsilon}^{A}\Gamma^{I}_{A\dot{A}}\gamma_{\mu}X^{I[a}\chi^{b]\dot{A}}
    \,, \quad
    \delta_\epsilon A_\mu^{aI} = i\,\bar{\epsilon}^{A}\Gamma^{I}_{A\dot{A}}\gamma_{\mu}\chi^{\dot{A}a}
    \,,
  \end{split}
 \end{equation}
provided that the linear constraints implied by \eqref{linglob} and the quadratic constraints \eqref{globalquad},
 are satisfied.

\renewcommand{\arraystretch}{1.5}

\begin{table}[ht]
\begin{center}
\begin{tabular}{||c||c|c|c|c||c||}
 \hline
 component & gauge vector & gauging & mass dim. & $V$ & interpretation\\
 \hline \hline
    $f_{abcd}$ & $A_\mu^{ab}$& $SO(N)$ & $(0,1)$ & $X^6$ & CS gauging \\[4pt]
    \hline
    $f_{abcI}$ & $\begin{array}{c} A_\mu^{aI} \\ A_\mu^{ab} \end{array}$
    & $\begin{array}{c} SO(N) \\ \mathbb{R}^{8N} \end{array}$ &
    $\begin{array}{c} (\ft12,\ft12) \\ (\ft12,1) \end{array}$ & $X^4$ & YM gauging \\[6pt]
    \hline
  $f^+_{IJKL}$& $-$& $-$ & $(1,-)$ & $X^2$ & massive CS \\[4pt]
 \hline
  $h_{IJ}$ & $A_{\mu}^{aI}$ & $\mathbb{R}^{8N}$& $(1,\ft12)$ & $X^2$ & top.~mass.~YM \\[4pt]
    \hline
\end{tabular}
\renewcommand{\arraystretch}{1}

\caption{\sl The different $SO(8)\times SO(N)$ representations of
the embedding tensor that survive the limit of global supersymmetry,
and the resulting gauging and gauge vectors (if applicable). The
next columns indicate the mass dimensions of the $\Theta_{\alpha
\beta}$ and $A_{\mu}{}^{\gamma}$ components and the order of the
resulting scalar potential. The interpretation of the different
models will be put forward in the next sections.}
\label{tab:gaugings}
\end{center}
\end{table}

To summarize, for any choice of the embedding tensor components
given in  \eqref{linglob} that satisfy  the quadratic constraints
\eqref{globalquad} we obtain a consistent globally supersymmetric
${\cal N}=8$ theory. The physical interpretation of the different
representations are quite different. We will illustrate this with a
few examples in the next section. For the moment we note that an
understanding of what these different representations signify can be
obtained from the covariant derivatives of the scalars and fermions
\eqref{cov-der-glob}. From these one can infer that $f_{abcd}$
induces a compact $SO(N)$ gauging, while $h_{IJ}$ gauges the
non-compact translations $\mathbb{R}^{8N}$. The representation
$f_{abcI}$ corresponds to a semi-direct product of compact and
non-compact gaugings in $SO(N) \ltimes \mathbb{R}^{8N}$. Finally,
the representation  $f_{IJKL}$ drops out from the covariant
derivatives and therefore is a massive deformation instead of a
gauging. For more information, see table \ref{tab:gaugings}.

Note that for our choice of scalings the R-symmetry $SO(8)$ is never
gauged; the components that give rise to the R-symmetry gaugings in
supergravity either drop out or are massive deformations.
Furthermore, from the table we conclude that only $f_{abcd}$
can give rise to a conformally invariant theory with a sextet
potential. In the next sections we will consider the various
representations separately.

\section{World-volume actions for multiple membranes} \setcounter{equation}{0}
In this section we consider different examples of globally
supersymmetric ${\cal N}=8$ theories obtained from gauged
supergravity in order to illustrate the different possible gaugings
outlined in the previous section (see the table). These can be
interpreted as different world-volume actions for multiple 2-branes.
An overview of our conventions can be found in appendix A.

\subsection{Conformal gaugings and multiple M2-branes}

In view of their applications to multiple M2-brane actions we first
consider the conformal gaugings with $f_{abcd}\ne 0$.

\subsubsection{Bagger-Lambert theory}

To reproduce the Bagger-Lambert theory we choose
\cite{Bergshoeff:2008cz}:
 \bea\label{BLtheta}
  \Theta_{ab,cd} \ = \ 2\, f_{abcd}\;,\hskip 3truecm f_{abcd}=f_{[abcd]}\,.
 \eea
This provides a particular solution of the linear constraints, while
the quadratic constraint reduces to the fundamental identity of
\cite{Bagger:2007jr}. The Lagrangian reads
 \begin{eqnarray}
  {\cal L} &=&
  \ft12D_{\mu}X^{Ia}D^{\mu}X^{Ia}-i\bar{\chi}^{\dot{A}a}\gamma^{\mu}D_{\mu}\chi^{\dot{A}a}
  + f_{abcd}\Gamma^{IJ}_{\dot{A}\dot{B}}X^{c}_I X^d_J
  \bar{\chi}^{\dot{A}a}\chi^{\dot{B}b} \\ \nonumber
  &&-\ft14\varepsilon^{\mu\nu\rho}f_{abcd}A_{\mu}{}^{ab}\left(\partial_{\nu}A_{\rho}{}^{cd}
  +\ft23 f^{d}{}_{efh}A_{\nu}{}^{ef}A_{\rho}{}^{ch}\right)
  -V\;,
 \end{eqnarray}
where the covariant derivatives are given by
 \bea\label{expressions}
   D_{\mu}X^{aI} \ = \
   \partial_{\mu}X^{aI}+A_ {\mu}{}^{cd}f_{cdab}X^{bI}\;, \qquad
   D_{\mu}\chi^{\dot{A}a} \ = \
   \partial_{\mu}\chi^{\dot{A}a}+A_{\mu}{}^{cd}f_{cdab}\chi^{\dot{A}b}\;.
 \eea
This is equivalent to the Bagger-Lambert action
\cite{Bagger:2007jr}. The supersymmetry variations (\ref{susy})
reduce to
 \begin{eqnarray}
  \delta_{\epsilon}X^{aI} &=&
  \bar{\epsilon}^{A}\Gamma^{I}_{A\dot{A}}\chi^{\dot{A}a}\;, \\
  \nonumber
  \delta_{\epsilon}\chi^{\dot{A}a} &=&
  \tfrac{i}{2}\Gamma^{I}_{A\dot{A}}\gamma^{\mu}\epsilon^{A}D_{\mu}X^{Ia}+
  \tfrac{1}{6}f_{abcd}\Gamma^{IJK}_{A\dot{A}}X^{b}{}_{I}X^{c}{}_{J}X^{d}{}_{K}\epsilon^{A}\;,
  \\ \nonumber
  \delta_{\epsilon}A_{\mu}{}^{ab} &=&
  -2i\bar{\epsilon}^{A}\Gamma^{I}_{A\dot{A}}\gamma_{\mu}X^{I[a}\chi^{b]\dot{A}}\;,
 \end{eqnarray}
in agreement with the superconformal symmetry of
\cite{Bagger:2007jr}.

\subsection{Non-conformal gaugings and multiple D2-branes}

We now consider the non-conformal gaugings with $f_{abcI}\ne
0$. As we will see, this representation  leads to supersymmetric Yang-Mills theories and multiple
D2-brane actions. We first discuss the non-semi-simple gaugings
triggered by this representation and next
consider the non-Abelian duality that converts the resulting action
into a supersymmetric Yang-Mills theory.

\subsubsection{Non-semi-simple gauge groups}

To construct  multiple D2-brane actions with kinetic Yang-Mills
terms we must consider gauge groups that are not semi-simple.
Specifically, this incorporates gauge groups, whose generators are
partially in the direction of the non-compact $t^{aI}$.

We consider the simplest case, where only $\Theta_{ab,cI}$ is
non-zero. The $SO(8)$ indices are decomposed according to $I=(i,8)$,
with $i=1,\ldots,7$, i.e. we are going to break the R-symmetry group
to $SO(7)$. The explicit ansatz is given by the completely
antisymmetric continuation of
 \bea\label{Lieansatz}
  \Theta_{ab,c 8} \ = \ - g_{\text{YM}}\,f_{abc}\;,
 \eea
while all other components vanish. Here, $f_{abc}$ are the structure
constants of an arbitrary $N$-dimensional Lie algebra with an
invariant tensor (in particular, $f_{abc}$ is totally antisymmetric)
and $g_{\text{YM}}$ is the Yang-Mills coupling constant which has
mass dimension $\ft12$. This ansatz gives rise to two types of gauge
group generators $X_{\alpha}$ according to (\ref{gaugegen}): either
proportional to $t^{ab}$ or $t^{aI}$. Denoting the former by ${\bf
X}$ and the latter by ${\bf T}$, respectively, this amounts to a
gauge algebra, which schematically reads
 \bea\label{semidirect}
  \big[{\bf X},{\bf X}\big] \ \subset \ {\bf X}\;, \qquad
  \big[{\bf X},{\bf T}\big] \ \subset \ {\bf T}\;, \qquad
  \big[{\bf T},{\bf T}\big] \ = \ 0 \;.
 \eea
More precisely, this describes a semi-direct product between, say, a
semi-simple Lie algebra $\frak{g}$ with structure constants
$f_{abc}$ and the ${\rm dim}(\frak{g})$ abelian translations ${\bf
T}$.

In order to verify that (\ref{Lieansatz}) gives rise to a consistent
gauging, we have to check the quadratic constraints. Following the
discussion in the previous section it turns out that the only
surviving quadratic constraint components are ${\cal Q}_{aI,bJ,cd}$
leading to the constraints
 \bea
  \Theta_{aI,c}{}^{g}\Theta_{gd,bJ}-\Theta_{aI,d}{}^{g}\Theta_{gc,bJ}
  -\Theta_{aI,gb}\Theta^{g}{}_{J,cd} \ = \ 0\;.
 \eea
For the ansatz (\ref{Lieansatz}), this is satisfied by virtue of the
Jacobi identities for $f_{abc}$, where we assume that its invariant
tensor is given by $\delta_{ab}$, possibly after a suitable change
of basis. We conclude that we can gauge an arbitrary Lie group.

\subsubsection{Multiple D2-branes through non-abelian duality}

The world-volume theories of multiple D2-branes are known to be
Yang-Mills gauge theories -- as opposed to the Chern-Simons gauge
theories discussed above -- and are not conformally invariant. In
fact, these two features are related, since in a non-abelian
Yang-Mills term the gauge coupling constant needs to be dimensionful
in $D\neq 4$, thus breaking the conformal invariance. In contrast,
in the Chern-Simons gauge theories the gauge coupling can be chosen
to be dimensionless.

To make contact with multiple D2-brane actions we now apply the
non-Abelian duality of \cite{Nicolai:2003bp} converting the
Yang-Mills Chern-Simons term into a standard Yang-Mills kinetic
term. We use the ansatz (\ref{Lieansatz}) for the embedding tensor,
where we may think of the structure constants as defining $SU(N)$.
There are two types of scalars, $X^{ai}\ (i=1,\ldots ,7)$ and
$\bar{X}^{a}=X^{a8}$, for which the covariant derivatives read
 \bea\label{shiftcov}
 \begin{split}
  D_{\mu}X^{ai} \ &= \
  \partial_{\mu}X^{ai}+g_{\text{YM}}f^{a}{}_{bc}A_{\mu}{}^{b}X^{ci}\;, \\
  D_{\mu}\bar{X}^{a} \ &= \ \partial_{\mu}\bar{X}^{a}+g_{\text{YM}}f^{a}{}_{bc}
  A_{\mu}{}^{b}\bar{X}^{c}+B_{\mu}{}^{a}\;,
 \end{split}
 \eea
where we defined
 \bea
  A_{\mu}{}^{a} \ \equiv \ A_{\mu}{}^{a8} \;, \qquad
  B_{\mu}{}^{a} \ \equiv \ \ft12\Theta_{a8,bc}A_{\mu}{}^{bc}\;.
 \eea
The resulting action reads
 \begin{eqnarray}\label{D2CS}
  {\cal L}_{\rm D2} &=&  \ft12 D_{\mu}X^{ia}D^{\mu}X^{ia}+\ft12
  D_{\mu}\bar{X}^{a}D^{\mu}\bar{X}^{a}
  -i\bar{\chi}^{\dot{A}a}\gamma^{\mu}D_{\mu}\chi^{\dot{A}a}\\ \nonumber
  &&-\ft12\varepsilon^{\mu\nu\rho}
  B_{\mu a}F_{\nu\rho}{}^{a}
  +\bar{\chi}^{a}f_{abc}\Gamma^{8i}X^{ib}\chi^{c}-V\;,
 \end{eqnarray}
with the non-abelian field strength
 \bea
  F_{\mu\nu}{}^{a}=\partial_{\mu}A_{\nu}{}^{a}-\partial_{\nu}A_{\mu}{}^{a}+ g_{\text{YM}}\,f^{a}{}_{bc}A_{\mu}{}^{b}A_{\nu}{}^{c}\;.
 \eea
The scalar potential is the quadratic expression in $A_2$, given by
 \bea
  A_{2}^{A,\dot{A}a} \ = \
  -\tfrac{1}{4}\Gamma^{ij8}_{A\dot{A}}g_{\text{YM}}\,f_{abc}X^{b}{}_{i}X^{c}{}_{j}\;.
 \eea
To see the equivalence to Yang-Mills gauge theories, we observe that
$B_{\mu}{}^{a}$ enters only algebraically and it can therefore be
integrated out. The St\"uckelberg shift symmetry on the extra
scalars $\bar{X}^{a}$ apparent in (\ref{shiftcov}) can be used to
gauge this scalar to zero. In turn, the field equations for
$B_{\mu}{}^{a}$ read
$B_{\mu}{}^{a}=\ft12\varepsilon_{\mu\nu\rho}F^{\nu\rho a}$. After
reinsertion into (\ref{D2CS}), we obtain a supersymmetric action
with Yang-Mills type kinetic term,
 \begin{eqnarray}
  {\cal L}_{\rm D2} &=&  \ft12
  D_{\mu}X^{ia}D^{\mu}X^{ia}-\tfrac{1}{4}F^{\mu\nu a}F_{\mu\nu}{}^{a}
  -i\bar{\chi}^{\dot{A}a}\gamma^{\mu}D_{\mu}\chi^{\dot{A}a}
  +\bar{\chi}^{a}f_{abc}\Gamma^{8i}X^{ib}\chi^{c}-V,
 \end{eqnarray}
which is the standard super-Yang-Mills action for D2-branes. This
dualization converted the topological gauge vectors into propagating
fields, then carrying the degrees of freedom of the scalars
$\bar{X}^{a}$.

It is instructive to compare our results on multiple D2-brane
actions with the recent proposal of
\cite{Gomis:2008uv,Benvenuti:2008bt,Ho:2008ei}.
 These theories contain two extra scalars  with
wrong-sign kinetic terms and thus may lead to ghosts. In a recent
development it has been pointed out that these ghosts can be avoided
if a different model is used where a translational symmetry, which
is present in the original theory, is gauged
\cite{Bandres:2008kj,Gomis:2008be,Ezhuthachan:2008ch}. After gauge
fixing the translational symmetry and integrating out some of the
fields one ends up with a  supersymmetric Yang-Mills theory. In this
context, we note that starting from three-dimensional Yang-Mills
theory the coupling constant $g_{\text{YM}}$ can be promoted to a
scalar field $X_+$ by replacing in the Lagrangian ${\cal
L}_{\text{YM}}$ the coupling constant $g_{\text{YM}}$ by $X_+$ and
adding to the Lagrangian a term with a Lagrange multiplier 2-form
gauge field $C_{\mu\nu}$ such that we obtain the total
Lagrangian\,\footnote{More generally, we may replace the full
embedding tensor by a set of scalar fields, see
\cite{Bergshoeff:2008cz}.}
\begin{equation}
{\cal L}_{\text{total}} = {\cal L}_{\text{YM}}  \ + \
\varepsilon^{\mu\nu\rho}\, \partial_\mu X_+\, C_{\nu\rho}\,.
\end{equation}
In a second step we define a vector field ${\tilde C}^\mu\equiv
\varepsilon^{\mu\nu\rho}\,C_{\nu\rho}$ and introduce a second scalar
field $X_-$ via the Stueckelberg redefinition ${\tilde C}^{ \mu} =
C^\mu -\partial^\mu X_-$ with the corresponding  shift symmetry
$\delta C^\mu = \partial^\mu\lambda\,, \delta X_- = \lambda$. This
leads to a gauge-equivalent Lagrangian of the form
\begin{equation}
{\cal L}_{\text{total}} = {\cal L}_{\text{YM}}  \ - \  \partial_\mu
X_+\bigl (\partial^\mu X_- - C^\mu\bigr)
\end{equation}
which is of the type  considered in
\cite{Bandres:2008kj,Gomis:2008be,Ezhuthachan:2008ch}.

\subsection{Massive deformations}

It is well-known that background fluxes may lead to massive
deformations of the worldvolume theory. Two exampes will be
discussed here: the first is triggered by a four-form flux in
M-theory and was recently considered in
\cite{Gomis:2008cv,Hosomichi:2008qk}, while the second is known to arise if
the mass parameter of IIA supergravity is turned on\cite{Bergshoeff:1996cy}. We will
show how these massive deformations also fall in the framework of
section \ref{general}.

\subsubsection{Massive Bagger-Lambert theory}

To reproduce the massive deformation of
\cite{Gomis:2008cv,Hosomichi:2008qk} we choose
\begin{eqnarray}
&& \Theta_{IJ,KL} = \mu\,\bigl(\epsilon_{\bar I \bar J \bar K \bar
L}\,, \epsilon_{\underaccent{\bar} I  \underaccent{\bar} J
\underaccent{\bar} K \underaccent{\bar} L}\bigr)\,, \qquad
\Theta_{ab,cd} = 2\, f_{abcd}\,,\qquad f_{abcd} = f_{[abcd]}\,,
\end{eqnarray}
where $\mu$ is a mass parameter and we have written the $SO(8)$
index $I$ as $I=(\bar{I}, \underaccent{\bar}{I})$ in terms of
$SO(4)\times SO(4)$ indices. The components $\Theta_{IJ,KL}$ are
self-dual, i.e. $\Theta_{IJ,KL}^-=0$, which is consistent with the
linear constraints. Note that the $SO(8)$ and $SO(N)$ sectors
decouple in the quadratic constraints.

With respect to the Bagger-Lambert theory the Yukawa coupling $A_2$
in \eqref{Yukawa-glob} contains an extra term proportional to $\mu$.
Since the potential is quadratic in $A_2$ we have  two further terms
in the potential: one mass term quadratic in $\mu$ and $X$  and one
flux term quartic in $X$ and linear in $\mu$. This precisely
reproduces the results of \cite{Gomis:2008cv, Hosomichi:2008qk}.


\subsubsection{Topologically massive D2-branes}

We next consider an embedding tensor given by the
symmetric-traceless tensor in \eqref{linconst} with respect to
$SO(8)$. There are several possible solutions as, for instance,
 \bea
  h_{88} \ = \ 1\;, \qquad
  h_{ij} \ = \ -\ft17 \delta_{ij}\;,
 \eea
where we have split again the indices according to $I=(i,8)$. This
breaks the R-symmetry to $SO(7)$. Therefore, it might be interpreted
as a D2 brane action in a massive IIA background, which is known to
lead to topologically massive vectors on the world-volume
\cite{Bergshoeff:1996cy}. Instead of constructing a specific model,
we are going to show that such a gauging generically leads to
topologically massive gauge vectors. The excited components are
$\Theta_{aI,bJ}$ and $\Theta_{IJ,KL}$. The latter component does not
appear in covariant derivatives, neither in Chern-Simons terms,
after consistently setting $A_{\mu}{}^{IJ}=0$. For the bosonic
couplings we therefore focus on the covariant derivatives
 \bea
  D_{\mu}\bar{X}^{aI} \ = \
  \partial_{\mu}\bar{X}^{aI}+m A_{\mu}{}^{aI}\;,
 \eea
where $A_{\mu}{}^{aI}=\Theta_{aI,bJ}A_{\mu}{}^{bJ}$ and we have
pulled out a mass parameter $m$, in accordance with the mass
dimension of $\Theta$. In the limit, the Chern-Simons term reduces
to an abelian term, such that one finds in total
 \bea
  {\cal L}_{\rm mD2} \ = \ \ft12
  D_{\mu}X^{aI}D^{\mu}X^{aI}
  -\ft12m\varepsilon^{\mu\nu\rho}A_{\mu}{}^{aI}\partial_{\nu}A_{\rho}{}^{aI}+\cdots\;,
 \eea
where we focused on the relevant couplings. It would be interesting
the inspect the supersymmetry in more detail. However, due to the
provision expressed in footnote 2, we postpone this to later work.
Using the shift symmetry, we can gauge-fix $X^{aI}$ to zero, after
which the equations of motion for the gauge vectors read
 \bea\label{masspin1}
  \varepsilon^{\mu\nu\rho}\partial_{\nu}A_{\rho}{}^{aI} \ = \
  mA^{\mu aI}\;.
 \eea
This describes one massive spin-1 degree of freedom
\cite{Townsend:1983xs}, which follows from the fact that
(\ref{masspin1}) implies the two equations
 \bea
  \partial^{\mu}F_{\mu\nu}{}^{aI} \ = \
  -\ft12m\varepsilon_{\nu\rho\lambda}F^{\rho\lambda aI} \ = \
  -m^2A_{\nu}{}^{aI}\;, \qquad
  \partial^{\mu}A_{\mu}{}^{aI} \ = \ 0\;.
 \eea
An equivalent description of a single massive degree of freedom
carried by a vector is given by the sum of a Maxwell term and a
Chern-Simons term. Both provide a gauge-invariant description of
massive vectors, which is a peculiarity in three dimensions.
Furthermore, it can be checked that the quadratic constraints allow
for the inclusion of $\Theta_{ab,cI}$ in addition to the symmetric
traceless component. With the Ansatz \eqref{Lieansatz} the
interpretation of this combination is clear: this is equivalent to a
topologically massive Yang-Mills theory.

\section{Conclusions} \setcounter{equation}{0}

In this work we derived a general framework for constructing gaugings and massive
deformations of ${\cal N} = 8$ (conformal and non-conformal) supersymmetric
gauge theories that describe
multiple membranes. Our starting point was gauged ${\cal N}=8$
supergravity in three dimensions. Performing the limit of global
supersymmetry and making different choices for the embedding tensor
we were able to reproduce a variety of membrane actions.

In particular, we have shown that the conformal gaugings, triggered
by the anti-symmetric $SO(N)$ representation $f_{abcd}$, led to the
Bagger-Lambert theory describing conformal invariant multiple
M2-brane actions. The non-conformal gaugings, triggered by the
$f_{abcI}$ representation, led to multiple D2-brane actions.
Finally, the self-dual anti-symmetric $SO(8)$ representation
$f^+_{IJKL}$ (in combination with $f_{abcd}$) led to a massive
deformation of the Bagger-Lambert theory, whereas the symmetric
traceless $h_{IJ}$ representation  led to a topologically massive
gauge theory.

We would like to stress that in addition to these known results, our
results also allow one to combine the different ingredients (subject
to the quadratic constraints). This would lead to generalisations of
the previously discussed theories, whose membrane interpretation
might be worth investigating. It is also worthwile to investigate whether
the procedure we introduced to define the limit of global supersymmetry
is unique or whether other limits are possible.

On a different note, in this paper we showed how starting from a
gauged supergravity theory a variety of globally supersymmetric
theories could be constructed. It would be interesting to apply this
technique to other situations as well. For instance, one could
consider cases with less supersymmetry and compare with the results
of \cite{Gaiotto:2008sd, Hosomichi:2008jd} for ${\cal N}=4$ and
\cite{Aharony:2008ug, Benna:2008zy} for ${\cal N}=6$. A
distinguishing difference between the ${\cal N}=8$ and ${\cal N}=6$
cases is that, whereas the conformal ${\cal N}=8$ embedding tensor
can only be defined as copies of the 4-index Levi-Civita symbol,
i.e., with $SO(4)$ gauge groups inside $SO(N)$ for $N=4k$ with $k$
integer, the conformal ${\cal N}=6$ embedding tensor can be defined
for any $U(N)$ gauge group \cite{deWit:2003ja}. This is related to
the fact that for ${\cal N}=6$ the global symmetry group is $U(N)$
and hence, using complex notation, the relevant embedding tensor can
be expressed in terms of Kronecker delta's, for any $N$, instead of
a Levi-Civita tensor, for special values of $N$. This fact has been
used in the recent constructions of \cite{Aharony:2008ug,
Benna:2008zy}.

Finally it would be interesting to apply the procedure to construct
globally supersymmetric theories out of locally supersymmetric theories
in different dimensions and, in particular, to see whether some of them
can be interpreted as the worldvolume theories of  multiple branes.

\subsection*{Acknowledgments}

We thank Bernard de Wit, Joaquim Gomis, Giuseppe Milanesi, Hermann
Nicolai, Teake Nutma, Henning Samtleben, and Stefan Vandoren for
useful discussions. This work was partially supported by the
European Commission FP6 program MRTN-CT-2004-005104EU and by the
INTAS Project 1000008-7928. In addition, the work of D.R.~has been
supported by MCYT FPA 2004-04582-C02-01 and CIRIT GC 2005SGR-00564.

\begin{appendix}
\renewcommand{\theequation}{\Alph{section}.\arabic{equation}}

\section{Useful relations} \setcounter{equation}{0}

\subsection{Conventions}

We use the following notation for the different indices:
 \begin{itemize}
 \item $I,J,\ldots = 1,\ldots,8$ for the $SO(8)$ R-symmetry vector indices, which will be split up according to:
  \begin{itemize}
  \item $I = (i,8)$ with $i,j,\ldots = 1,\ldots,7$ when the R-symmetry is broken to $SO(7)$,
  \item $I = (\bar{I}, \underaccent{\bar}{I})$ with $\bar I, \bar J\ldots = 1,\ldots,4$ and $\underaccent{\bar} I, \underaccent{\bar} J \ldots = 5,\ldots,8$ when the R-symmetry is broken to $SO(4) \times SO(4)$,
  \end{itemize}
 \item $A,B,\ldots=1,\ldots,8$ for the $SO(8)$ R-symmetry  spinor indices,
 \item $\dot{A},\dot{B},\ldots =1,\ldots,8$ for the $SO(8)$ R-symmetry conjugate spinor indices,
 \item $a,b,\ldots 1,\ldots,N$ for the $SO(N)$ fundamental indices,
 \item $\cI,\cJ,\ldots = (I,a)$ for the $SO(8,N)$ fundamental indices,
 \item $\alpha=[\cI\cJ]=([IJ],[ab],Ia)$ for the $SO(8,N)$ adjoint indices,
 \item $\underaccent{\bar}{\alpha},\underaccent{\bar}{\beta},\ldots$ are flat indices corresponding to the local $SO(8)\times SO(N)$ action.
\end{itemize}
Note that our conventions differ from those of
\cite{Bergshoeff:2008cz} by an $SO(8)$ triality rotation in order to
be compatible with the supergravity results of
\cite{Nicolai:2001ac}. Moreover, we employ the convention that
summation over the antisymmetric index pairs $[ab]$ and $[IJ]$ is
accompanied by a factor of $\ft12$.

\subsection{$SO(8,N)$ structures}

In order to compute the various components of the Maurer-Cartan
forms and of ${\cal V}^{\alpha}{}_{\underaccent{\bar}{\alpha}}$ to
lowest order as used in the main text, we need the explicit form of
the Lie algebra $\frak{so}(8,N)$. In covariant form it reads
 \bea
  \big[t^{\cI\cJ},t^{\cK\cL}\big] \ = \ 2\left(
  \eta^{\cI[\cK}t^{\cL]\cJ}-\eta^{\cJ[\cK}t^{\cL]\cI}\right)\;,
 \eea
with the indefinite $\eta^{\cI\cJ}=(\delta^{IJ},-\delta^{ab})$.
Splitting this in a $SO(8)\times SO(N)$ covariant form (and
redefining $t^{ab}\rightarrow -t^{ab}$) one finds
 \begin{eqnarray}\nonumber\label{so8N}
  \big[t^{IJ},t^{KL}\big] &=& 2\left(
  \delta^{I[K}t^{L]J}-\delta^{J[K}t^{L]I}\right)\;, \quad
  \big[t^{ab},t^{cd}\big] \ = \ 2\left(
  \delta^{a[c}t^{d]b}-\delta^{b[c}t^{d]a}\right)\;, \\
  \big[t^{IJ},t^{Ka}\big] &=& -2\delta^{K[I}t^{J]a}\;, \qquad
  \big[t^{ab},t^{Ic}\big] \ = \ -2\delta^{c[a}t^{Ib]}\;, \\
  \nonumber
  \big[t^{Ia},t^{Jb}\big] &=&
  \delta^{IJ}t^{ab}+\delta^{ab}t^{IJ}\;.
 \end{eqnarray}
This corresponds to the following structure constants
 \bea\label{strucconst}
 \begin{split}
  f^{ab,cd}{}_{ef} \ &= \
  8\delta^{[a}{}_{[e}\delta^{b][c}\delta^{d]}{}_{f]}\;, \qquad
  f^{IJ,KL}{}_{PQ} \ = \
  8\delta^{[I}{}_{[P}\delta^{J][K}\delta^{L]}{}_{Q]}\;, \\
  f^{IJ,Ka}{}_{Lb} \ &= \
  -2\delta^{K[I}\delta^{J]}{}_{L}\delta^{a}{}_{b}\;, \qquad
  f^{ab,cI}{}_{dJ} \ = \
  -2\delta^{c[a}\delta^{b]}{}_{d}\delta^{I}{}_{J}\;, \\
  f^{Ia,Jb}{}_{cd} \ &= \
  \delta^{IJ}\delta^{a}{}_{[c}\delta^{b}{}_{d]}\;,
 \end{split}
 \eea
where we use the conventions that summation over antisymmetric
indices is accompanied by a factor of $\ft12$, i.e.,
$[t^{ab},t^{cd}]=\ft12 f^{ab,cd}{}_{ef}t^{ef}$, etc.

For the computations of ${\cal
V}^{\alpha}{}_{\underaccent{\bar}{\alpha}}$ one has to insert
(\ref{scalar}) into (\ref{adjoint}) and use the first of the BCH
relations
 \bea\label{BCH}
 \begin{split}
  e^{-A}Be^{A} \ &= \ B+\big[B,A\big] +\tfrac{1}{2!}\big[
  [B,A],A\big]+\tfrac{1}{3!}\big[ [[B,A],A],A\big]+\cdots\;, \\
  e^{-A}de^{A} \ &= \ dA+\tfrac{1}{2!}\big[ dA,A]+\tfrac{1}{3!}\big[
  [dA,A],A\big]+\cdots\;.
 \end{split}
 \eea
This yields the following components
 \begin{equation}\label{Vcomp}
  \begin{split}
   {\cal V}^{ab}{}_{cd}\ &= \ 2\delta^{a}{}_{[c}\delta^{b}{}_{d]}
   +2\delta^{[a}{}_{[c}X^{b]I}X_{d]}^{I}+{\cal O}(X^{4})\;, \\
   {\cal V}^{ab}{}_{cI} \ &= \ 2\delta^{[a}{}_{c}X^{b]}{}_{I}
   -X^{[a}{}_{I}X^{b]J}X^{J}{}_{c}-\ft13
   X^{[a}{}_{J}\delta^{b]}{}_{c}X^{Jd}X^{d}{}_{I}
   +\cO (X^4)\;, \\
   {\cal V}^{ab}{}_{IJ} \ &= \ -2 X^{[a}{}_{I}X^{b]}{}_{J}+\cO
   (X^4)\;, \\
   \cV^{IJ}{}_{ab} \ &= \
   -2X^{I}{}_{[a}X^{J}{}_{b]}+\cO(X^4)\;, \\
   \cV^{IJ}{}_{KL} \ &= \
   2\delta^{I}{}_{[K}\delta^{J}{}_{L]}+2\delta^{[I}{}_{[K}X^{J]a}X^{a}{}_{L]}+\cO(X^4)\;,
   \\
   \cV^{IJ}{}_{Ka} \ &= \
   -2X^{[I}{}_{a}\delta^{J]}{}_{K}
   -X^{[I}{}_{a}X^{J]b}X^{b}{}_{K}-\ft13
   X^{L}{}_{a}X^{Lb}X^{b[I}\delta^{J]}{}_{K}
   +\cO(X^4)\;, \\
   \cV^{Ia}{}_{bc} \ &= \ -2 X^{I}{}_{[b}\delta^{a}{}_{c]}
   -\ft23 X^{I}{}_{[b}X^{J}{}_{c]}X^{Ja}
   +\cO(X^4)\;,
   \\
   \cV^{Ia}{}_{Jb} \ &= \
   \delta^{I}{}_{J}\delta^{a}{}_{b}+2X^{a[I}X^{J]}{}_{b}+\cO(X^4)\;,
   \\
   \cV^{Ia}{}_{JK} \ &= \ -2X^{a}{}_{[J}\delta^{I}{}_{K]}
   -X^{Ib}X^{a}{}_{[J}X^{b}{}_{K]}
   +\cO(X^4)\;.
  \end{split}
 \end{equation}
For the various components of the Maurer-Cartan forms (\ref{maurer})
one finds by use of the second of the BCH formulas (\ref{BCH})
 \bea
 \begin{split}\label{PQ}
  \cQ_{\mu}{}^{ab} \ &= \
  \partial_{\mu}X^{I[a}X^{b]I}+\Theta_{\alpha\beta}A_{\mu}{}^{\alpha}\cV^{\beta}{}_{ab}+\cO(X^3)\;, \\
  \cQ_{\mu}{}^{IJ} \ &= \ \partial_{\mu}X^{a[I}X^{J]a}+
  \Theta_{\alpha\beta}A_{\mu}{}^{\alpha}\cV^{\beta}{}_{IJ}+\cO(X^3)\;, \\
  \cP_{\mu}{}^{Ia} \ &= \ D_{\mu}X^{Ia}+\cO(X^3)\;,
  \label{P-Q}
 \end{split}
 \eea
where the covariant derivative reads
 \bea
  D_{\mu}X^{aI} \ = \ \partial_{\mu}X^{aI}+\Theta_{\alpha\beta}A_{\mu}{}^{\alpha}\cV^{\beta}{}_{Ia}\;.
  \label{cov-der}
 \eea
Furthermore we expand the T-tensor:
 \begin{eqnarray}\label{TExpan}\nonumber
  T_{IJ,KL} &=& -\Theta_{ab,IJ}X^{a}{}_{K}X^{b}{}_{L}
  +4\Theta_{aI,bK}X^{a}{}_{J}X^{b}{}_{L} 
  +2\Theta_{aI,KL}X^{a}{}_{J}+([IJ]\leftrightarrow [KL])\big)\\
  &&+\Theta_{IJ,KL}
  +\cO
  (X^3)\;, \\ \nonumber
  T_{IJ,Ka} &=&
  -\Theta_{ab,cd}X^{b}{}_{K}X^{c}{}_{I}X^{d}{}_{J}+\Theta_{bc,LK}X^{b}{}_{I}X^{c}{}_{J}X^{L}{}_{a}\\
  \nonumber
  &&-\ft12\Theta_{IJ,bc}X^{b}{}_{K}X^{cL}X^{L}{}_{a}
  +\ft16\Theta_{IJ,ab}X^{Lb}X^{Ld}X^{d}{}_{K}\\ \nonumber
  &&-\ft12\Theta_{IJ,PQ}X^{P}{}_{a}X^{Qb}X^{b}{}_{K}
  +\ft16\Theta_{IJ,KL}X^{P}{}_{a}X^{Pb}X^{Lb} \\ \nonumber
  &&-\Theta_{aK,bc}X^{b}{}_{I}X^{c}{}_{J}+2\Theta_{ab,c[I}X^{c}{}_{J]}X^{b}{}_{K}+2\Theta_{KL,b[I}X^{b}{}_{J]}X^{L}{}_{a}
  \\ \nonumber
  &&+2\Theta_{aK,b[I}X^{b}{}_{J]}+ \Theta_{IJ,ab}X^{b}{}_{K}+\Theta_{IJ,KL}X^{L}{}_{a}+\Theta_{IJ,Ka}+\cO
  (X^4)\;, \\ \nonumber
  T_{IJ,ab} &=&
  -\Theta_{ab,cd}X^{c}{}_{I}X^{d}{}_{J}-\Theta_{IJ,KL}X^{K}{}_{a}X^{L}{}_{b}
  \\ \nonumber
  &&+2\Theta_{ab,c[I}X^{c}{}_{J]}+2\Theta_{IJ,K[a}X^{K}{}_{b]}
  -4X^{c}{}_{[I}\Theta_{J]c,K[a}X^{K}{}_{b]}+\Theta_{IJ,ab}+\cO
  (X^3)\;,
 \end{eqnarray}
where we suppressed in the first line an antisymmetrization in
$[IJ]$ and $[KL]$.

\end{appendix}

\end{document}